\documentclass[aps,prl,twocolumn]{revtex4}
\usepackage{amsmath,bm,epsfig}

\def\Fbox#1{\vskip1ex\hbox to 8.5cm{\hfil\fboxsep0.3cm\fbox{%
  \parbox{8.0cm}{#1}}\hfil}\vskip1ex\noindent}  

\newcommand{\B}[1]{{\bm{#1}}}
\newcommand{\C}[1]{{\mathcal{#1}}}    
\let \= \equiv \let\*\cdot \let\~\widetilde \let\^\widehat \let\-\overline

\def\<{\left\langle}    \def\>{\right\rangle}
\def\({\left(}          \def\){\right)}
 \def \[ {\left [} \def \] {\right ]}

\begin{document}
\title{Statistical Mechanics and Dynamics of a 3-Dimensional Glass-Forming System}
\author{Edan Lerner, Itamar Procaccia and Jacques Zylberg}
\affiliation{$^1$Department of Chemical Physics, The Weizmann Institute of Science, Rehovot 76100, Israel}
\
\date{\today}

 \begin{abstract}
In the context of a classical example of glass-formation in 3-dimensions we exemplify how to construct a statistical mechanical theory of the glass transition. At the heart of the approach is a simple criterion for verifying a proper choice of up-scaled quasi-species that allow the construction of a theory with a finite number of 'states'. Once constructed, the theory identifies a typical scale $\xi$ that increases rapidly with lowering the temperature and which determines the $\alpha$-relaxation time $\tau_\alpha$ as $\tau_\alpha \sim \exp(\mu\xi/T)$ with $\mu$ a typical chemical potential. The theory can predict relaxation times at temperatures that are inaccessible to numerical simulations.
 \end{abstract}
  \maketitle

{\bf Introduction}: Among the best studied models of the glass transitions are those employing point-particles with a soft binary potential.
Some repeatedly studied examples are the Kobb-Andersen model \cite{93KA}, the Shintani-Tanaka model \cite{06ST}, the
Dzugutov model \cite{89Dzu} and various versions of binary mixtures with purely repulsive potentials, see for example \cite{89DAY,99PH,08BPGSD}. While easy to simulate on the computer, these models are challenging for theorists due to the fact that it is extremely hard to evaluate statistical-mechanical partition-function integrals in continuous coordinates. It is therefore very tempting to find a reasonable up-scaling (coarse-graining) method that would define a discrete statistical-mechanics with partition sums rather than integrals, with
the sum running on a finite number of quasi-species which have well characterized degeneracies and enthalpies. Indeed, in a number of examples in 2-dimensions it was shown that such a discrete statistical-mechanics is possible \cite{07ABIMPS,07HIMPS,07ILLP,08LP,09LPR} and quite advantageous \cite{08HIP,08HIPS} in providing a successful description of the statistics and the dynamics of systems undergoing the glass transition. In this Letter
we offer a general criterion for the selection of up-scaled quasi-species and demonstrate it, for the first time, in the context
of a 3-dimensional model system undergoing a glass transition.

{\bf Model}: We employ here a version of a much studied model consisting of a 50:50 mixture of $N$ point-particles in 3-dimensions ($N=4096$ in our case), interacting via a binary potential. We refer to half the particles as 'small' and half as `large'; they interact via a pairwise potential $U(r_{ij})$:
\begin{equation}
\label{potential}
U(r_{ij}) =
\left\{
\begin{array}{ccl}
\epsilon\left[ \left(\frac{\sigma_{ij}}{r_{ij}} \right)^\alpha
- \left(\frac{\sigma_{ij}}{r_{ij}} \right)^\beta + a_0 \right] & , & r_{ij} \le r_c(i,j) \\
0 & , & r_{ij} > r_c(i,j)
\end{array}
\right.
\end{equation}
Here, $\epsilon$ is the energy scale and $\sigma_{ij} = 1.0, 1.2$ or 1.4 for small-small, small-large
or large-large interactions, respectively. For the sake of numerical speed the potential is cut-off smoothly at a distance,
denoted as $r_c$, which is calculated by solving
$\partial U/\partial r_{ij}|_{r_{ij} = r_c} = 0$ which translates to $r_c =
\left({\alpha}/{\beta}\right)^{\frac{1}{\alpha-\beta}}\sigma_{ij}$. The parameter $a_0$ is chosen to guarantee
the condition $U(r_c) = 0$. Below we use $\alpha=8$ and $\beta=6$, resulting in
$r_c = \sqrt{8/6}~\sigma_{ij}$ and $a_0 = 0.10546875$.

\begin{figure}
\hskip -1.2cm
\centering
\includegraphics[scale = 0.45]{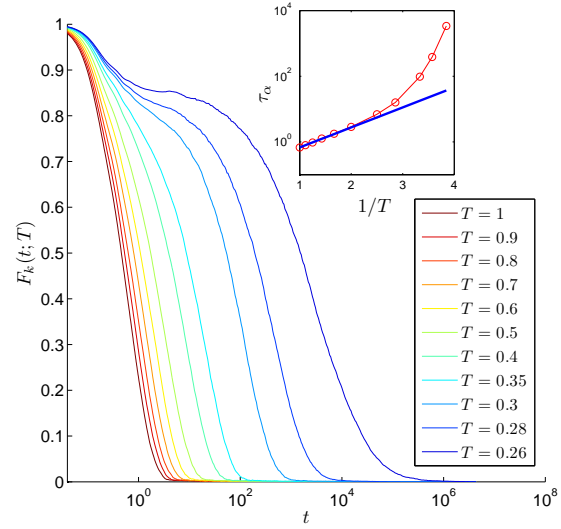}
\caption{Color online: Time dependence of the correlation functions (\ref{defFk}) for a range of temperatures (decreasing from left to right) as shown in the figure. The inset shows the relaxation time $\tau_\alpha$ in a log-lin plot vs $1/T$, compared to an Arrhenius temperature dependence.}
\label{corr}
\end{figure}

As in many of these models, one can quantify the slowing down in the super-cooled regime by measuring a typical correlation function.
Here we measured the self-part of the intermediate scattering function \cite{99PH} summed over the large particles only,
\begin{equation}
F_k(t;T) \equiv \left\langle \case{2}{N}\sum_{i=1}^{N/2} \exp\left\{{i\B k \cdot [\B r_i(t)-\B r_i(0)]}\right\}\right \rangle  \ . \label{defFk}
\end{equation}
In Fig. \ref{corr} we show these correlation functions for $k=5.1\sigma^{-1}$ and for a range of temperatures as indicated in the figure. We see the usual rapid slowing down that can be measured by introducing the typical time scale $\tau_\alpha$ that is determined by noting the time where $F_k(t=\tau_\alpha;T)= F_k(0;T)/e \equiv 1/e$. The relaxation times are shown in the inset of Fig. \ref{corr} as a function of $1/T$ in a log-lin plot to stress the non-Arrhenius dependence at lower temperatures.

\begin{figure}
\hskip -2.0cm
\includegraphics[scale = 0.435]{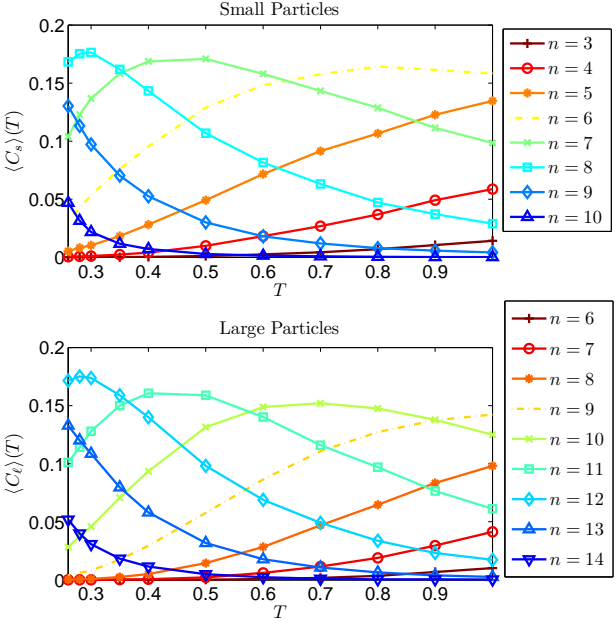}
\caption{Color online: Temperature dependence of the concentrations of the various quasi-species. Symbols are simulation data and the lines are a guide to the eye. }
\label{conc}
\end{figure}

{\bf Statistical Mechanics}: Our aim is to provide a statistical mechanical theory that captures the structural changes upon lowering the temperature such that there will pop-up a typical scale that can be used to predict the relaxation time $\tau_\alpha$. To this aim we need to up-scale (coarse-grain) from particles to quasi-species that can be characterized by their enthalpy and degeneracy.  Up-scaling can be done in various ways and there is no unique algorithm to select a-priori a `best' up-scaling. Here we offer a criterion to validate a chosen up-scaling. We choose to work
with particles and their nearest neighbors, where `neighbors' are defined as all the particles $j$ around a chosen central particle $i$ that are within the range of interaction $r_c(i,j)$. In the interesting range of temperatures we find 8 quasi-species with one `small' central particle and $3,4\dots 10$ neighbors, and 9 quasi-species with one `large' central particle with $6,7\dots 14$ neighbors, all in all 17 quasi-species. Other combinations have negligible concentration ($<0.5\%$) throughout the temperature range. We denote these
quasi-species as $C_s(n)$ and $C_\ell(n)$ with $s$ and $\ell$ denoting the small or large central particle, while $n$ denotes the number of neighbors. We measured the mole-fractions $\langle C_s(n)\rangle(T)$ and $\langle C_\ell(n)\rangle(T)$ and the results are shown in Fig. \ref{conc}.
\begin{figure}
\hskip -1.0cm
\centering
\includegraphics[scale = 0.50]{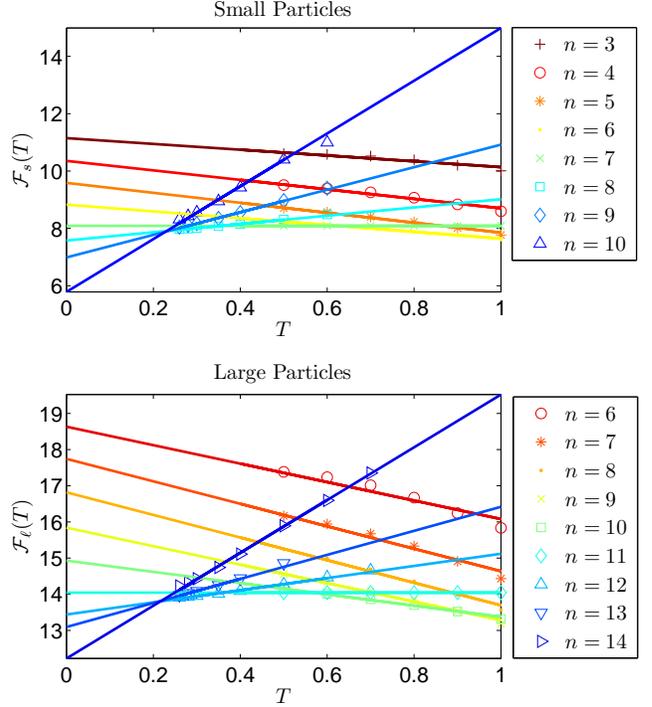}
\caption{Color online: The approximate linear dependence of the free energies of the chosen quasi-species on the temperature.
From the slope we read the degeneracy and from the intercept the enthalpies (up to normalization), cf. Eq. \ref{readHg}. Note that when the free energies are large we do not have data: the concentrations become exponentially small and in a finite simulation box they disappear completely.}
\label{linear}
\end{figure}

\begin{figure}
\hskip -0.6 cm
\centering
\includegraphics[scale = 0.55]{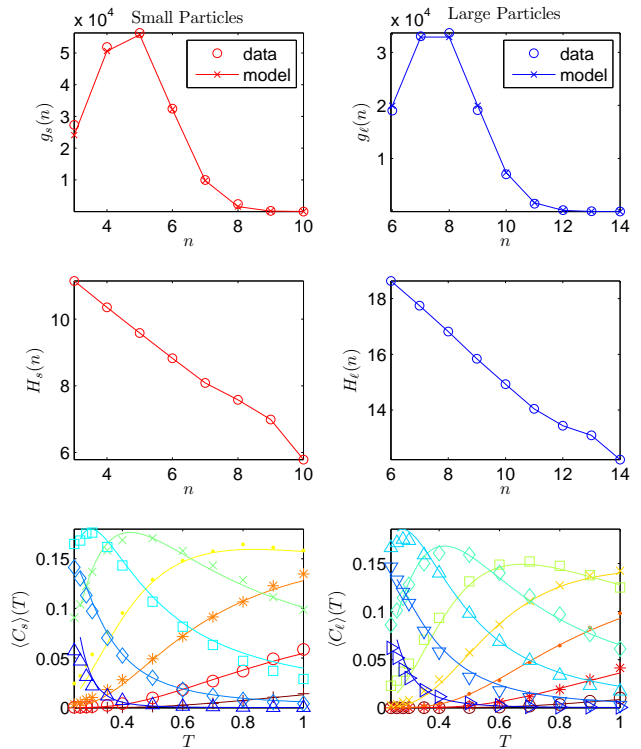}
\caption{Color online: Upper panel: The degeneracies $g_s(n)$ and $g_\ell(n)$ read from the slopes of Fig. \ref{linear} (in circles) and the degeneracies according to the gaussian model Eq. (\ref{comparisons}). Middle panel: the measured enthalpies. Lower panel: comparison of the measured concentrations of quasi-species to those calculated from Eqs. (\ref{DefgH}) using the model degeneracies and measured enthalpies. Here symbols are data and lines are theoretical predictions.}
\label{comparisons}
\end{figure}

To decide whether this up-scaling provides a useful statistical mechanics we now ask whether there exist
free energies $\C F_s(n;T)$ and $\C F_\ell(n;T)$ such that
\begin{eqnarray}
\langle C_s(n)\rangle (T)& =& \frac{ e^{-\C F_s(n;T)/T}}{2\sum_{n=3}^{10}  e^{-\C F_s(n;T)/T}} \ ,\nonumber\\
\langle C_\ell(n)\rangle (T)& =& \frac{e^{-\C F_\ell(n;T)/T}}{2\sum_{n=6}^{14} e^{-\C F_\ell(n;T)/T}} \ . \label{DefF}
\end{eqnarray}
The free energies are found by inverting Eqs. (\ref{DefF}) in terms of the measured concentrations. We then plot these quantities
as a function of the temperature, as demonstrated for the present case in Fig. \ref{linear}. If $\C F_s(n;T)$ and $\C F_\ell(n;T)$ can be well approximated as linear in the temperature, we can interpret
\begin{eqnarray}
\C F_s(n;T) &\equiv& H_s(n) - T \ln g_s(n)\nonumber \ , \nonumber\\
\C F_\ell(n;T) &\equiv& H_\ell(n) - T \ln g_\ell(n) \ , \label{readHg}
\end{eqnarray}
where now the degeneracies $g_s(n)$ and $g_\ell(n)$ (read from the slopes in Fig. \ref{linear} and enthalpies $H_s(n)$ and $H_\ell(n)$ (read from the
intercepts) are {\bf temperature-independent}. This validates the choice of up-scaling. In other words, the approximate linearity of the inverted free energies in the temperature means that we can write the concentrations as
\begin{eqnarray}
\langle C_s(n)\rangle (T)& \approx& \frac{g_s(n) e^{-H_s(n)/T}}{2\sum_{n=3}^{10} g_s(n) e^{-H_s(n)/T}} \ ,\nonumber\\
\langle C_\ell(n)\rangle (T)&\approx& \frac{g_\ell(n) e^{-H_\ell(n)/T}}{2\sum_{n=6}^{14} g_\ell(n) e^{-H_\ell(n)/T}} \ . \label{DefgH}
\end{eqnarray}
Then we can use these forms also as a prediction for temperatures where the simulation time is too short to observe the relaxation. The resulting degeneracies $g_s(n)$ and $g_\ell(n)$ can be easily modeled theoretically, given basically by a Gaussian distribution around the most probable number $n_{\rm mp}$ of nearest neighbors for small and large particles respectively:
\begin{eqnarray}
g_s(n)&\approx& e^{-[(n-n^s_{\rm mp})^2/2\sigma^2_s ]} \ , \quad n^s_{\rm mp}=4.65, \sigma^2_s=1.55\ , \nonumber\\
g_\ell(n)&\approx& e^{-[(n-n^\ell_{\rm mp})^2/2\sigma^2_\ell ]}\ , \quad n^\ell_{\rm mp}=7.50,  \sigma^2_\ell=2.0 \ . \label{gasussian}
\end{eqnarray}
 The comparison of the theoretical to the measured degeneracies
is shown in Fig. \ref{comparisons}, upper panel. The same figure shows in the middle panel the enthalpies of the various quasi-species.
One could model the enthalpies as a linear function in $n$.
 These results are easily interpreted; we have high enthalpies when there are large free volumes (few neighbors). The lowest enthalpies are found when there are many neighbors and there is no much costly free volume. In other words, at the present density and range of temperatures the $pV$ term dominates the energy in the enthalpy.  Using the theoretical degeneracies and the measured enthalpies we compute the concentrations of all our quasi-species and compare them with the measurement in the lowest panel of Fig. \ref{comparisons}. The agreement that we have, especially considering the number of quasi-species and the simplicity of the theory, is very satisfactory. Notice that the competition between degeneracy and enthalpy explains the rather intricate temperature-dependence of the concentrations of the quasi-species, sometimes declining when the temperature drops, sometime rising, and sometime having non-monotonic behavior.

\begin{figure}
\hskip -1.6 cm
\centering
\includegraphics[scale = 0.45]{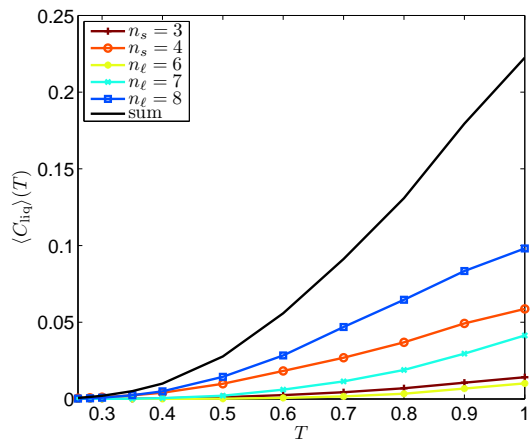}
\caption{Color online: The temperature dependence of $C_{\rm liq}(T)$ is shown as the upper continuous line. The contributions of the
various `liquid' sub-species are shown with symbols which are identified in the inset.}
\label{cliq}
\end{figure}

{\bf Prediction of Relaxation Times}: Finally, we want to connect the structural theory to the dynamical slowing-down. To this aim we note that there are a number
of quasi-species whose concentration goes down exponentially (or maybe faster) when the temperature decreases, and that the
relaxation time shoots up at the same temperature range. We refer to these quasi-species as the `liquid' ones; in this example the liquid concentrations are those with small particles with three and four neighbors, and large particles with six, seven and eight neighbors. We sum up these concentrations and denote the sum as $\langle C_{\rm liq}\rangle (T)$. The dependence of $\langle C_{\rm liq}\rangle (T)$ on the temperature is shown in Fig. \ref{cliq}

This concentration is used to define
a typical scale $\xi(T)$,
\begin{equation}
\xi(T) \equiv [\rho C_{\rm liq}(T)]^{-1/3} \ ;
\end{equation}
where $\rho$ is the number density. This length scale has the physical interpretation of the average distance between the `liquid' quasi-species. It was argued before \cite{07ILLP,08LP,08EP} that this length scale can be also interpreted as the linear size of relaxation events which include $O(\xi(T))$ quasi-species. We can therefore estimate the growing free energy per relaxation event as $\Delta G=\mu \xi(T)$ where $\mu$ is the typical chemical
potential per involved quasi-species. This estimate, in turn, determines the relaxation time as
\begin{equation}
\tau_\alpha(T)= e^{\mu\xi(T)/T} \ . \label{fit}
\end{equation}
\begin{figure}
\hskip -1.2cm
\centering
\includegraphics[scale = 0.37]{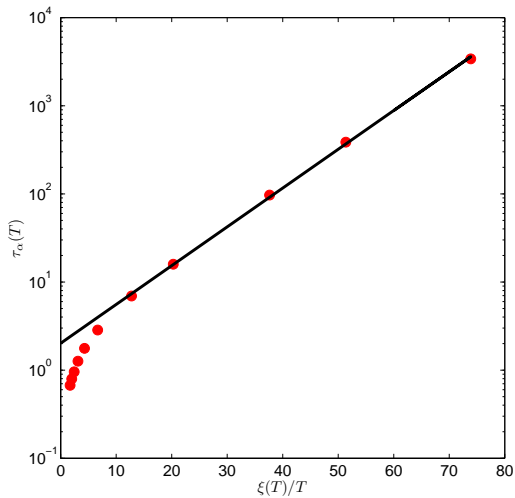}
\caption{Color online: The relaxation time $\tau_\alpha(T)$ in terms of the typical scale $\xi(T)$. We show the excellent fit to
Eq. (\ref{fit}) with $\mu=0.04$. Note that the intercept at $T\to \infty$ is of the order of unity as it must be.}
\label{tauvsxi}
\end{figure}
The quality of this prediction can be gleaned from Fig. \ref{tauvsxi}, where we can see that the fit is excellent, with $\mu\approx 0.04$. The fact that the intercept in Fig. \ref{tauvsxi} is of the order of unity is very reassuring, since this is what we expect
when $T\to \infty$.

A few points should be stressed. As we expect (cf. Ref \cite{08EP}), in systems with point particles and soft potential, there is no reason to
fit the relaxation time to a Vogel-Fulcher form \cite{96EAN} which predicts a singularity at finite temperature. In our approach
we predict that $\xi\to \infty$ only when $T\to 0$, and there is nothing singular on the way, only slower and slower relaxation. At some
point the simulation time will be too short for the system to relax, but we can use Eq. (\ref{fit}) to predict what should be the simulation time to allow the system to reach equilibrium.

{\bf Conclusions}: In summary, we reiterate that point particles with soft potential are different from granular media or systems of hard spheres which can truly jam and lose ergodicity. Point particles with soft potential remain ergodic \cite{08EP}, and therefore should be amenable
in their super-cooled regime to statistical mechanics. To construct simple, workable statistical mechanics one needs to up-scale the system and find a collection of quasi-species with well defined enthalpies and degeneracies. In this Letter we proposed a simple criterion to validate a choice of up-scaling, and demonstrated how, once the structural theory is under control, a natural length scale appears and can be used to determine the relaxation time, also for temperatures that cannot be simulated due to the fast growth of necessary relaxation time. The fact that the present approach works equally well in two and three dimensions provides good reason to
believe that it has a substantial degree of generality. How to use this kind of theory to understand relaxation functions and
thermodynamic properties was demonstrated before in two-dimensional examples, but there is still a large variety of related problem
in two and three dimensions that can profit from this approach.

\end{document}